\documentclass[
 reprint,
superscriptaddress,
showkeys,
 amsmath,amssymb,
 aps,
]{revtex4-2}

\usepackage{graphicx}
\usepackage{dcolumn}
\usepackage{bm}
\usepackage{hyperref}
\usepackage{xcolor}
\usepackage{titlecaps}

\begin{document}

\preprint{APS/123-QED}

\title{Active Dynamics of Linear Chains and Rings in Porous Media}

\author{Ligesh Theeyancheri}
\affiliation{Department of Chemistry, Indian Institute of Technology Bombay, Mumbai 400076, India}
\author{Subhasish Chaki}  
\affiliation{Department of Chemistry, Indian Institute of Technology Bombay, Mumbai 400076, India}
\affiliation{Department of Materials Science and Engineering, University of Illinois Urbana-Champaign, Urbana, Illinois 61801, USA}
\author{Tapomoy Bhattacharjee}
\email{tapa@ncbs.res.in}
\affiliation{National Centre for Biological Sciences, Tata Institute of Fundamental Research, Bangalore 560065, India}
\author{Rajarshi Chakrabarti}
\email{rajarshi@chem.iitb.ac.in}
\affiliation{Department of Chemistry, Indian Institute of Technology Bombay, Mumbai 400076, India}

\begin{abstract}
\noindent To understand the dynamical and conformational properties of deformable active agents in porous media, we computationally investigate the dynamics of linear chains and rings made of active Brownian monomers. In porous media, flexible linear chains and rings always migrate smoothly and undergo activity-induced swelling. However, semiflexible linear chains though navigate smoothly, shrink at lower activities, followed by swelling at higher activities, while semiflexible rings exhibit a contrasting behavior. Semiflexible rings shrink, get trapped at lower activities, and escape at higher activities. This demonstrates how activity and topology interplay and control the structure and dynamics of linear chains and rings in porous media. We envision that our study will shed light on understanding the mode of transport of shape-changing active agents in porous media. 
\end{abstract}


\maketitle
\section{Introduction}\label{Intro}
\noindent A class of active agents lives in complex and heterogeneous porous environments like gels, tissues, soils, and sediments~\cite{ribet2015bacterial,datta2016polymers,wu2005signaling}. For example, the microorganisms like bacteria move through the disordered environment in search for nutrients~\cite{wu2005signaling,bhattacharjee2021,liu2021viscoelastic}, natural killer cells scan through porous tissues to neutralize diseased cells~\cite{moretta2002natural,daher2018next,ferlazzo2012natural}, and biopolymers like motor proteins move through the living cells by ciliary and flagellar motility~\cite{brangwynne2008cytoplasmic,lau2003microrheology,sens2020stick}. In addition, bio-engineered polymers have been used in targeted drug delivery. The physiological barriers in the kidneys have a nanoporous structure that regulates the filtration of these drug carriers. These biological swimmers and their artificial analogs, such as bio-synthetic polymers, experience different types of confinements and interactions while navigating through porous environments~\cite{hess2001molecular,goel2008harnessing,heeremans2022chromatographic}. Depending on the topology, they often switch the modes of migration by deforming their shapes through their natural habitat to efficiently explore the medium for fulfilling their needs~\cite{lohrmann2023optimal}. Thus the deformability and topology of active agents bring additional complexity, which in turn either facilitates or suppresses their transport in complex media.\\

\noindent In simple liquid media, the inherent nonequilibrium nature of the active particles enables them to exhibit transient superdiffusion followed by a long-time enhanced diffusion~\cite{}. However, novel non-equilibrium effects emerge in the conformational and dynamical properties of a chain of interlinked active particles. Flexible chains swell with increasing activity~\cite{goswami2022reconfiguration,osmanovic2017dynamics,shin2015facilitation,samanta2016chain,chaki2019enhanced}. In contrast, semiflexible polymers shrink at low activity and swell at large activity~\cite{eisenstecken2016conformational}. Numerical simulations reported a coil-to-globule-like transition of active polymer chain due to the interplay of activity and thermal fluctuations~\cite{bianco2018globulelike}. Over the past few years, experimental and theoretical studies have focused on how the motion of the active agents in disordered media is influenced by the interactions with the obstacles in the neighborhood~\cite{volpe2011microswimmers,shin2015facilitation,theeyancheri2022migration,chopra2022geometric, mokhtari2019dynamics,irani2022dynamics,bhattacharjee20183d,kurzthaler2021geometric,moore2023active,lohrmann2023optimal,theeyancheri2022silico}. Recent studies have shown that the micro-confinement of the porous medium dramatically alters the run and tumble motion of rod-shaped bacterial cells to hopping and trapping motion~\cite{bhattacharjee2019bacterial}. Chopra $et.$ $al.$ experimentally studied the dynamics of non-tumbling \textit{E coli} bacteria in a square array of micropillars and reported an anomalous size-dependent active transport in the two-dimensional structured porous environments, where shorter cells get trapped while longer ones escape through the channel-like space between the pillars~\cite{chopra2022geometric}. Computer simulations of linear active polymers in the two-dimensional periodic porous medium demonstrated that the stiff chains are able to move almost unhindered through the ordered porous medium, whereas the flexible one gets stuck~\cite{mokhtari2019dynamics}. A recent study predicted that the local geometry determines the optimal path of the active agents by controlling the reorientations in locally dilute and dense regions~\cite{irani2022dynamics}. Moreover, recent studies on circular topology described the structure, dynamics, and emergence of new topological states of the active ring polymers~\cite{mousavi2019active,philipps2022dynamics,chubak2020emergence}. However, less attention has been devoted to active circular polymeric agents in the porous environment, despite their relevance in biological systems i.e, bacterial or mitochondrial DNA in eukaryotic cells, extruded loops in chromatin, and actomyosin rings~\cite{goloborodko2016chromosome,sehring2015assembly,gupta2021role}. The dynamics is even more complex and intriguing in porous media, where activity and topology of active agents interplay. Earlier experimental studies reported the role of polymer topology in pharmacokinetics for treating cancer. They found that the ring topology offers better performance than linear polymers due to reduced renal filtration and longer blood circulation time~\cite{fox2009soluble,chen2009influence}. Thus, the topology of the active agents is a crucial factor for modulating their biophysical properties and biological performance~\cite{scholz2014comb,nasongkla2009dependence,kwok2011comparative}. However, the physics behind the interplay between activity, confinement, and topology of active agents has not been well understood till date. Hence, a systematic and detailed investigation of the integrated effects of topology and conformations of active agents in complex porous media is of prime importance. \\
\begin{figure*}
\centering
\includegraphics[width=0.7\linewidth]{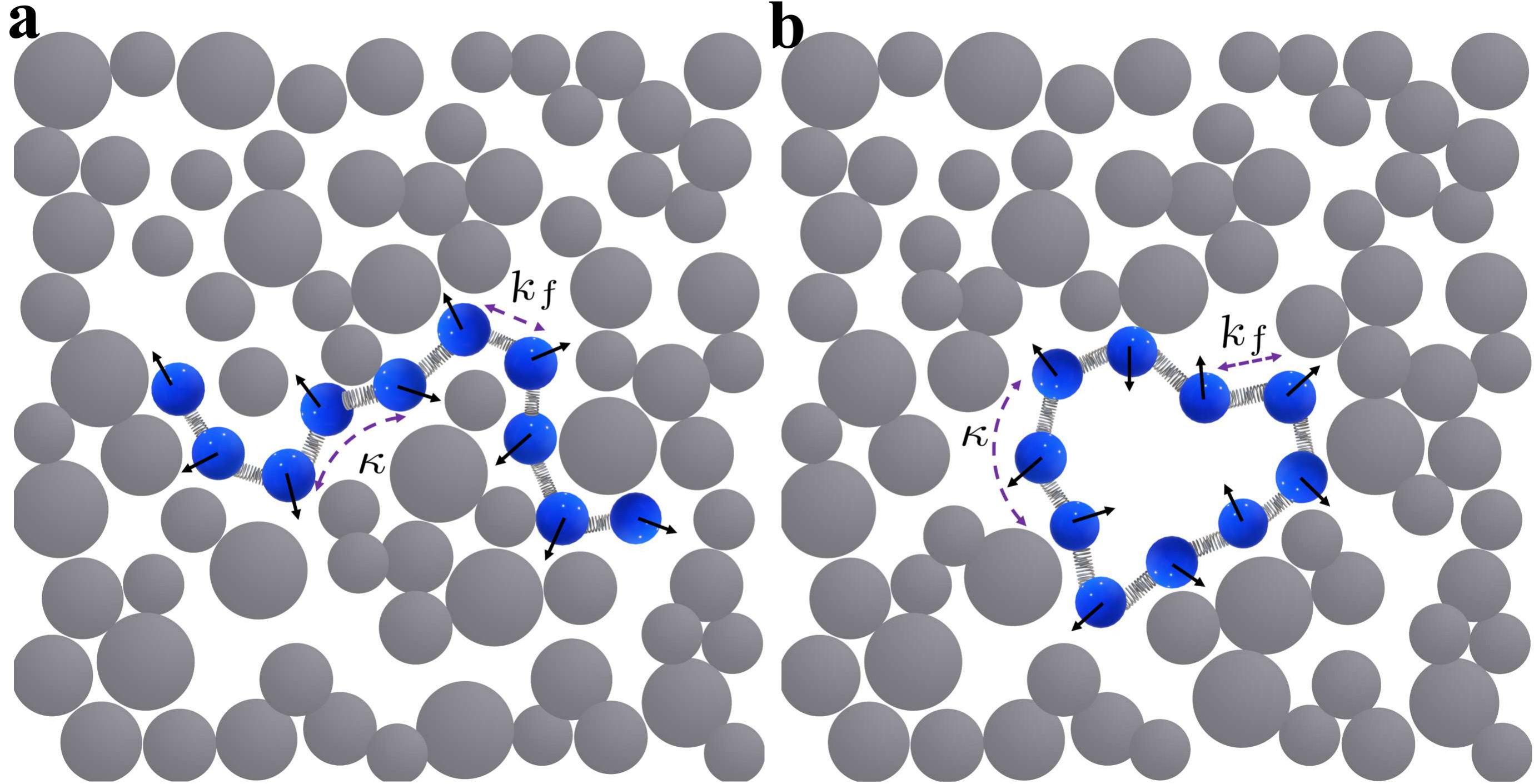} \\
\caption{\small The schematic sketch of the active (a) linear chain and (b) ring in porous media. The black arrows represent the instantaneous direction of the active force on the monomers. The linear chain occupies multiple pores by adapting straight conformations, which is not possible for the ring due to the geometrical constraint.}\label{fig:model}
\end{figure*}

\noindent In order to understand the mode of migration adopted by deformable active agents in crowded media, we computationally analyze the dynamical and conformational properties of active linear chains and rings made of active Brownian particles in two-dimensional static porous media. We consider three different types of polymers by changing their spring constant ($k_f $) and bending rigidity ($\kappa$) as flexible, inextensible, and semiflexible linear chains and rings. For both the flexible and inextensible polymers, the bending rigidity is zero. However, the inextensible polymers have a very large spring constant value than the flexible polymers. For semiflexible polymers, we chose a very high bending rigidity value with a spring constant the same as that of flexible. We find that the dynamics of both the linear chains and rings in porous media are enhanced owing to the mutual contribution of activity and the conformational fluctuations. The flexible or inextensible linear chains migrate faster compared to the rings made of the same number of monomers in porous media. In contrast, a reverse trend in dynamics is observed in unconfined space. i.e., linear chains move slowly in comparison to the rings in unconfined space. Semiflexible linear chains also move smoothly through the porous media, while semiflexible rings transiently get trapped in the pore confinements and escape from such traps with increasing activity. Conformational fluctuations of the flexible, inextensible, or semiflexible linear chains and rings exhibit contrasting behavior in the presence of activity. In porous media, flexible linear chains and rings exhibit activity-induced swelling irrespective of the fact that they are topologically distinct. In contrast, inextensible linear chains and rings display activity-induced shrinking. Surprisingly, semiflexible linear chains and rings behave differently in porous media depending on their topology. Semiflexible linear chains exhibit activity-induced swelling, whereas the rings show activity-induced shrinking with increasing activity. Our study reports how the combined effects of activity, confinements, and topological constraints facilitate or control the transport of active linear chains and rings in complex environments. 

\section{Method}\label{Model}
\subsection{Model and Simulation Details}

\noindent We model the disordered random porous media by randomly placing $M$ number (M = 1200, 2000, 2500) of static obstacles that are allowed to overlap inside a 2D square box of fixed size 300 $\sigma$. The size of the beads forming the porous medium, $\sigma_P$ ranges from 1 to 5 $\sigma$, and the size distribution of these particles falls under a Gaussian distribution with mean $\sim 3$ $\sigma$. The linear chain is modeled as a sequence of N self-propelled beads of diameter $\sigma$ connected by N-1 finitely extensible springs (\textbf{Fig.~\ref{fig:model}a}). The ring with the same number of active Brownian particles (monomers) is created by connecting the terminal beads of the linear chain by the same finitely extensible spring (\textbf{Fig.~\ref{fig:model}b}). In our simulation, $\sigma$, $k_B T$, and $\tau=\frac{\sigma^2 \gamma}{k_B T}$ set the units of length, energy, and time scales, respectively. Where $k_B$ is the Boltzmann constant, $T$ is the temperature, and $\gamma$ is the friction coefficient. The equation of motion for the linear chains and the rings has the general expression given by the following Langevin equation with an additional term to account for the activity. Here, we consider a high friction limit. Therefore, the dynamics is practically overdamped as the contribution from the inertia term is negligible, and hence we do not write the inertia term in the following equation of motion.
\begin{equation}
\gamma \frac{d \textbf{r}_{i}}{dt} = - \sum_{j} \nabla V(\textbf{r}_i-\textbf{r}_j) + {\bf f}_{i}(t) + {\bf{F}_{\text{a, i}}(t)}
\label{eq:langevineq_c4}
\end{equation}
here the drag force, $\gamma \frac{d \textbf{r}_{i}}{dt}$ is the velocity of $i^{th}$ bead times the friction coefficient $\gamma$, $r_i$ (i = 1, 2, ..., N) is the positions of the monomers of the linear chain or ring, $V(r)$ is the resultant pair potential between $i^\text{th}$ and $j^\text{th}$ particles accounting for the conservative forces, thermal force $\bf f_{i}(t)$ is modeled as Gaussian white noise with zero mean and variance $\left<f_{i}(t^{\prime})f_{j}(t^{\prime\prime})\right> = 4 \gamma k_B T \delta_{ij}\delta(t^{\prime}-t^{\prime\prime})$, and ${\bf{F}_{\text{a, i}}(t)}$ is the active force which drives the system out of equilibrium. ${\bf{F}_{\text{a, i}}(t)}$ has the magnitude $\text{F}_{\text{a}}$, acts along the unit vector~\cite{bechinger2016active} of each $i^{th}$ monomer, $\bf{n}(\boldsymbol{\theta_i}) = {(\textrm{cos} \,\boldsymbol{\theta_i}, \, \textrm{sin} \, \boldsymbol{\theta_i})}$, where $\theta_i$ evolves as $\frac{d \boldsymbol{\theta_i}}{dt} = \sqrt{2D_R} {\bf f}_{i}^{R}$, $D_R$ is the rotational diffusion coefficient and ${\bf f}_{i}^{R}$ is the Gaussian random number with a zero mean and unit variance. Hence, the persistence time of the individual monomers is related to the rotational diffusion coefficient, $D_R$ as $\tau_R = \frac{1}{D_R}$. The activity can also be expressed in terms of a dimensionless quantity, i.e., the P\'eclet number Pe, which is defined as $\frac{\text{F}_a \sigma}{k_BT}$. The total potential energy of the linear chains or rings can be written as $V(r) = V_{\text{FENE}} + V_{\text{BEND}} + V_{\text{WCA}}$ consists of bond, bending and excluded volume contributions. The bond stretching is controlled by the FENE potential:
\begin{equation}
V_{\text{FENE}}\left(r_{ij}\right)=\begin{cases} -\frac{k_f r_{\text{max}}^2}{2} \ln\left[1-\left( {\frac{r_{ij}}{r_{\text{max}}}}\right) ^2 \right],\hspace{3mm} \mbox{if } r_{ij} \leq r_{\text{max}}\\
\infty, \hspace{35mm} \mbox{otherwise}.
\end{cases}
\label{eq:FENE_c4}
\end{equation}
where $r_{ij}$ is the distance between two neighboring monomers in the linear chain or ring with a maximum extension of $r_{\text{max}} = 1.5 \sigma$, and $k_f$ is the spring constant~\cite{kremer1990dynamics}. To impose the condition of inextensibility, $k_f$ is set to be very high for the inextensible linear chains and rings ($k_f = 1000$). The stiffness of the linear chains and rings is implemented through the bending potential,
\begin{equation}
V_{\text{BEND}}\left(\phi_i \right) = \kappa \left(1-\cos\phi_i\right)
\label{eq:BEND_c4}
\end{equation}
where $\kappa$ is the bending modulus and $\phi_{i}$ is the angle between the bond vectors $i$ and $i+1$. To account for self-avoidance, a pair of monomers of the linear chains or rings interact $via$ the repulsive Weeks–Chandler–Andersen (WCA) potential~\cite{weeks1971role}.
\begin{equation}
V_{\textrm{WCA}}(r_{ij})=\begin{cases}4\epsilon_{ij} \left[\left(\frac{\sigma_{ij}}{r_{ij}}\right)^{12}-\left(\frac{\sigma_{ij}}{r_{ij}}\right)^{6}\right]+\epsilon_{ij}, \mbox{if }r_{ij}<2^{1/6}\sigma_{ij} \\
0, \hspace{35mm} \mbox{otherwise},
\end{cases}
\label{eq:WCA_c4}
\end{equation}
where $r_{ij}$ is the separation between the interacting particles, $\epsilon_{ij} = 1$ is the strength of the steric repulsion, and $\sigma_{ij} = \frac{\sigma_i + \sigma_j}{2}$ determines the effective interaction diameter, with $\sigma_{i(j)}$ being the diameter of the interacting pairs. The static obstacles in the porous media also interact repulsively with the monomers of linear chains or rings. We consider three different cases: flexible ($k_f = 30$ and $\kappa = 0$), inextensible ($k_f = 1000$ and $\kappa = 0$), and semiflexible ($k_f = 30$ and $\kappa = 1000$) linear chains and rings in the porous media.\\

\noindent All the simulations are performed using the Langevin thermostat, and the equation of motion is integrated using the velocity Verlet algorithm in each time step. We initialize the system by randomly placing the linear chains or rings inside the porous media and relaxing the initial configuration for $2 \times 10^6$ steps. All the production simulations are carried out for $10^9$ steps where the integration time step is considered to be $10^{-5}$, and the positions of the monomers are recorded every $100$ step. The simulations are carried out using Large-scale Atomic/Molecular Massively Parallel Simulator (LAMMPS)~\cite{plimpton1995fast}, a freely available open-source molecular dynamics package.

\subsection{Characterization of Porous Media}
\noindent We characterize the pore space structure by the average pore size $\xi$. We place passive tracer particles at different random locations in the media and allow them to diffuse through the polydisperse porous media. The average pore size $\xi$ is calculated from the time-and-ensemble averaged mean square displacements (MSD) of the passive tracer particle in the porous medium (Fig.~S1a). At longer time, the tracer gets confined by the obstacles, and the MSD saturates. Then we take the square root of this saturated value and add the tracer particle diameter to get $\xi$. The values of $\xi$ obtained from different tracer trajectories are binned to construct a histogram from which the ensemble-averaged probability distribution of $\xi$, $\text{P}(\xi)$ is computed (Fig.~S1b). To change the average pore size, $\xi$, we increase the obstacle density by adding particles, keeping the same width of the Gaussian distribution of the sizes of the obstacles. The average pore sizes of these different porous media are either comparable or smaller/larger than the average $\text{R}_g$ ($\text{R}_g^{0}$) of the passive linear chains or rings in unconfined space.

\section{Results and Discussion}\label{results}

\noindent We first simulate flexible linear chains and rings of different sizes (N = 20, 50, 80, 100, 150, 200) in unconfined space to validate our model. According to Flory's theory, $R_g$ of a polymer chain in good solvent scales as $N^\nu$, where $\nu = 0.75$ in 2D. We report that the exponent $\nu$ for the linear chain and ring are $0.76\pm 0.008$ and $0.74\pm 0.0072$, respectively (Fig.~S2). This implies that our model is consistent with the scaling predictions~\cite{rubinstein2003}. \\

\noindent To quantify the dynamical behavior of the active linear chains and rings, we compute the time-and-ensemble averaged Mean Squared Displacement (MSD) of the center of mass (COM) ($\text{r}_c$) of the linear chains or rings, $\left\langle{\overline{\Delta \text{r}_{c}^{2}(\tau)}}\right\rangle$, and scaling exponent, $\alpha(\tau)=\frac{d \log \left<\overline{\Delta \text{r}_c^{2}(\tau)}\right>}{d \log \tau}$ as a function of lag time $\tau$. We consider the chain length as $N = 50$ and vary the activity, $\text{F}_{a}$. In the unconfined case, $\left\langle{\overline{\Delta \text{r}_{c}^{2}(\tau)}}\right\rangle$ of the flexible, inextensible, and semiflexible active linear chains and rings display a three-step growth: a short time thermal diffusion, intermediate superdiffusion, and enhanced diffusion at longer times as compared to the overdamped dynamics of the passive case~\cite{wu2014three,eisenstecken2017internal,bhattacharjee20183d,philipps2022dynamics} (Fig.~S3). Polymer topology results the linear chains to move slower than the rings in the unconfined space due to their larger size (Fig.~S3).

\subsection{Navigation of Linear Chains and Rings in Porous Media: Smooth Migration, Trapping, and Escaping}
\begin{figure*}
\centering
\includegraphics[width=0.9\linewidth]{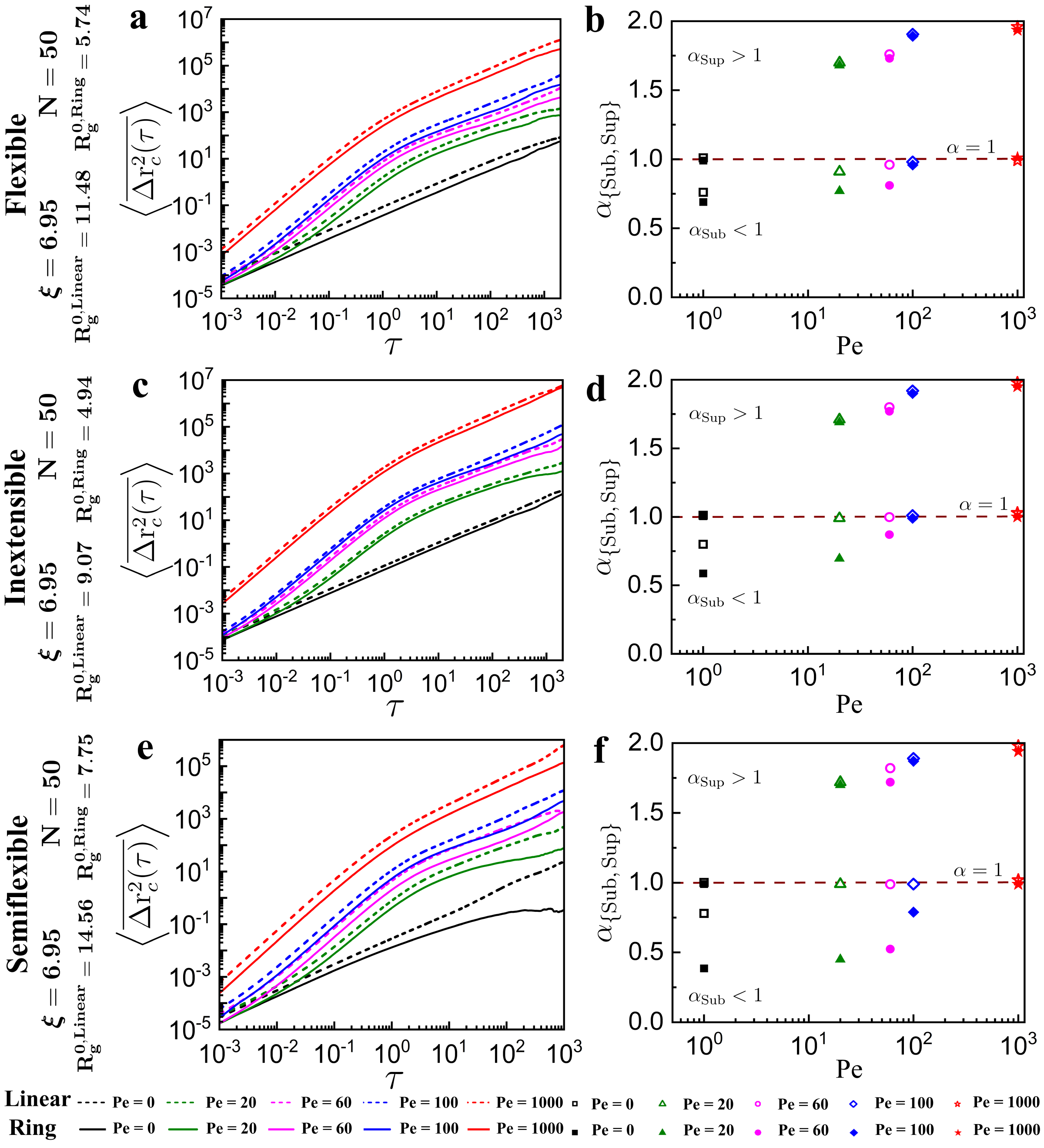}
\caption{\small $\left\langle{\overline{\Delta \text{r}_{c}^{2}(\tau)}}\right\rangle$ $vs$ $\tau$ and $\alpha_{\{\text{Sub, Sup}\}}$ $vs$ $\text{Pe}$ for flexible (a, b), inextensible (c, d), and semiflexible (e, f) linear chains (dashed lines or open symbols) and rings (solid lines or solid symbols) (N = 50) subjected to different activity in the porous medium with $\xi = 6.95$. $\alpha_{\{\text{Sub, Sup}\}}$ illustrates the subdiffusion ($\alpha_{\text{Sub}} < 1$) and superdiffusion ($\alpha_{\text{Sup}} > 1$) exponents (obtained from $\alpha(\tau)$ in Fig.~S4), respectively. Brown dashed line represents $\alpha = 1$.}\label{fig:dynamics}
\end{figure*} 

\begin{figure*} 
\centering
\includegraphics[width=0.85\linewidth]{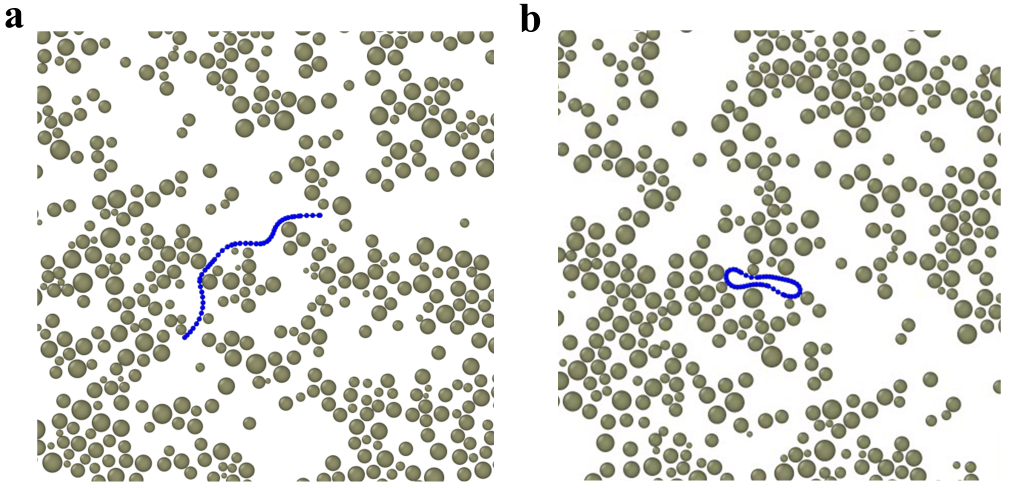} 
\caption{\small The snapshots of active ($\text{Pe} = 60$) semiflexible (a) linear chain and (b) ring in the porous medium with $\xi = 6.95$. The semiflexible linear chain smoothly moves while the ring gets trapped inside the pore confinement. The snapshots are zoomed in for better clarity (see Movie S5 and Movie S6). }\label{fig:snapshot}
\end{figure*}

\noindent In porous media, the dynamics is controlled by pore confinements in addition to the activity and polymer topology, which may lead to different behaviors as compared to in unconfined space. We consider active linear chains and rings in porous media with $\xi = 6.95$. Irrespective of the topologies the flexible or inextensible linear chains and rings navigate smoothly through the pores undergoing conformational fluctuations (Movie S1-S4). Although there exist different regions of motion: intermediate superdiffusion and long-time enhanced diffusion  due to activity  (Fig.~S4), a noticeable difference with the unconfined space is that the linear chains now migrate faster than the rings made of the same number of monomers (\textbf{Fig.~\ref{fig:dynamics}}). This designates the imperative role of polymer topology in crowded media. However, the dynamics differ dramatically for the semiflexible linear chains from the rings in porous media. The semiflexible linear chains move smoothly by preferentially adopting straight/rod-like conformations (\textbf{Fig.~\ref{fig:snapshot}a} and Movie S5) and occupying multiple pores simultaneously (Fig.~S5), while the micro-confinements can trap and restrict the motion of the semiflexible rings in the porous media (\textbf{Fig.~\ref{fig:snapshot}b} and Movie S6), due to their stretched circular conformations (Fig.~S5). As a result of the trapping, the subdiffusive behavior of the semiflexible ring is more pronounced and persists for a longer time than the semiflexible linear chain at lower activities (\textbf{Fig.~\ref{fig:dynamics}(e, f)}). An increase in activity enhances the conformational fluctuations of semiflexible rings, which facilitate their escape from the pore confinements (\textbf{Fig.~\ref{fig:dynamics}}). Hence, our analyses manifest that the topology of the active agent greatly influences and regulates their transport in the porous media.    

\subsection{Effect of Porous Architecture on Migration of Active Linear Chains and Rings}
\begin{figure*}
\centering
\includegraphics[width=0.8\linewidth]{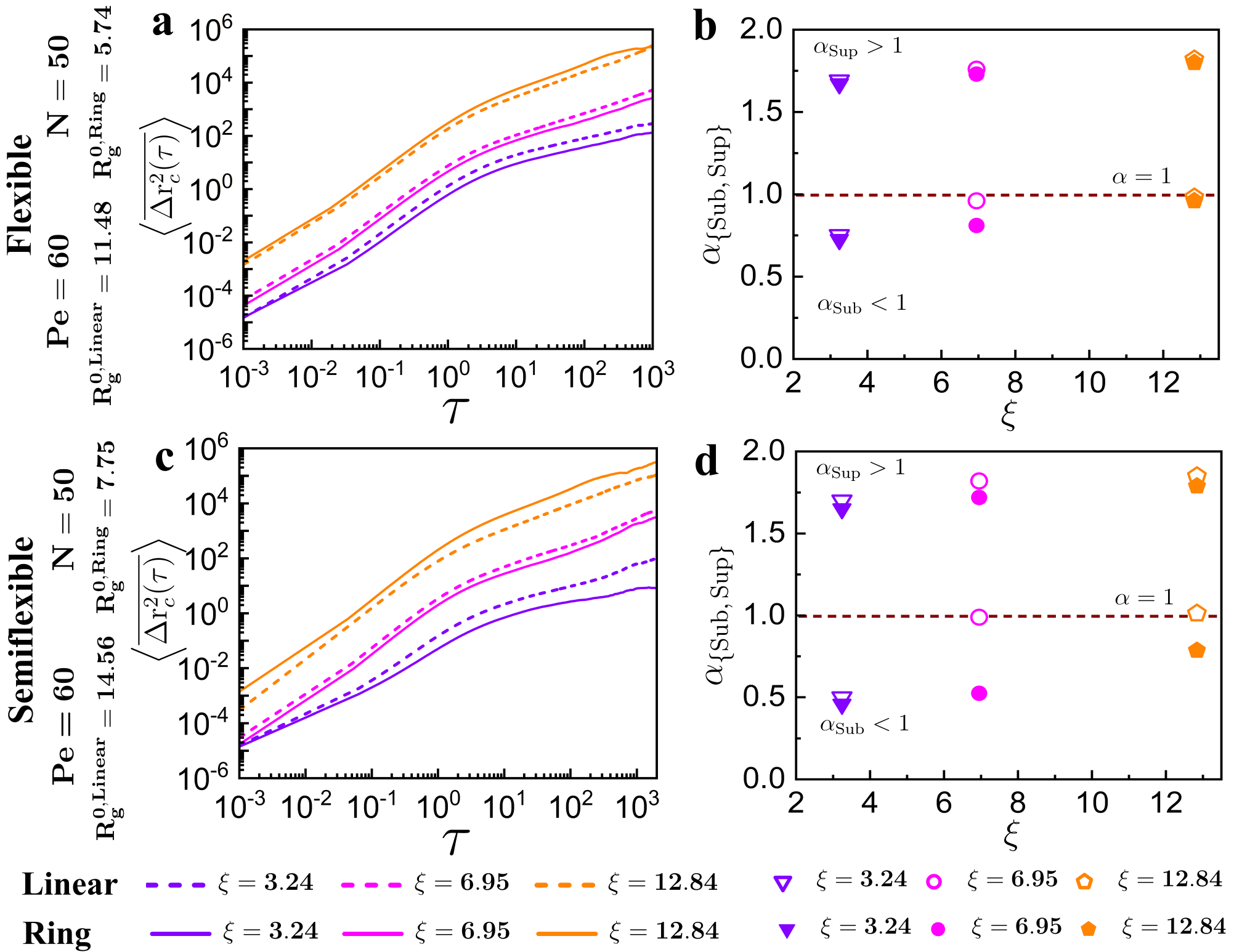}
\caption{\small $\left\langle{\overline{\Delta \text{r}_{c}^{2}(\tau)}}\right\rangle$ $vs$ $\tau$ and $\alpha_{\{\text{Sub, Sup}\}}$ $vs$ $\xi$ for active ($\text{Pe} = 60$) flexible (a, b), and semiflexible (c, d) linear chains (dashed lines or open symbols) and rings (solid lines or solid symbols) (N = 50) in different porous media. $\alpha_{\{\text{Sub, Sup}\}}$ illustrate the subdiffusion ($\alpha_{\text{Sub}} < 1$) and superdiffusion ($\alpha_{\text{Sup}} > 1$) exponents (obtained from $\alpha(\tau)$), respectively. Brown dashed line represents $\alpha = 1$.}\label{fig:dynamics_media} 
\end{figure*}

\noindent Further, to characterize how the local arrangement of the media affects the transport of linear chains and rings, we simulate flexible and semiflexible linear chains and rings in porous media with different average pore sizes ($\xi$) for a given activity $\text{Pe} = 60$. If the size of the ring is comparable or larger than $\xi$, the flexible and semiflexible linear chains migrate faster than the rings (\textbf{Fig.~\ref{fig:dynamics_media}}). A reverse trend in dynamics is observed like in unconfined media if the average size of the ring is much smaller than $\xi$, but the dynamics becomes comparatively slower in porous media than in the unconfined space. For very small $\xi$, the semiflexible active linear chains still migrate smoothly, while the semiflexible active rings get trapped indicated by strong subdiffusive behavior, and become diffusive at larger $\xi$ (\textbf{Fig.~\ref{fig:dynamics_media}} and Fig.~S6). There is a larger difference in the dynamics of the semiflexible linear chains and rings as their motion is significantly affected by the pore confinements.

\subsection{Conformations of Linear Chains and Rings in Unconfined and Porous Media: Topology Dependent Activity-induced Swelling and Shrinking} 

\begin{figure*}
\centering
\includegraphics[width=0.8\linewidth]{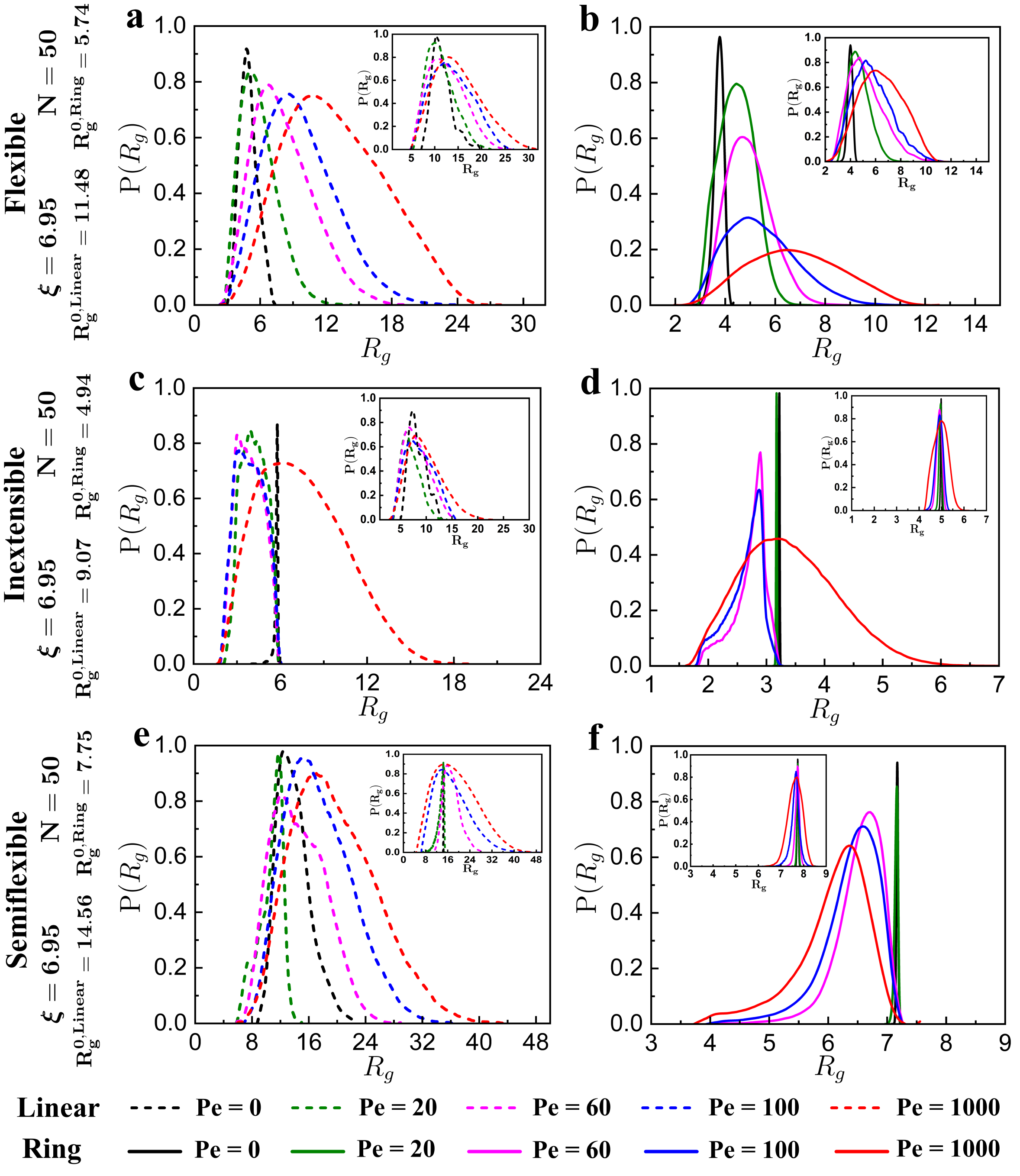}
\caption{$\text{P}(R_g)$ $vs$ $R_g$ for flexible (a, b), inextensible (c, d), and semiflexible (e, f) linear chains (dashed lines) and rings (solid lines) subjected to different activity in the porous medium with $\xi = 6.95$. The insets represent the same for the respective linear chains and rings in unconfined space subjected to different $\text{Pe}$.}\label{fig:conformation}
\end{figure*}

\begin{figure*}
\centering
\includegraphics[width=0.85\linewidth]{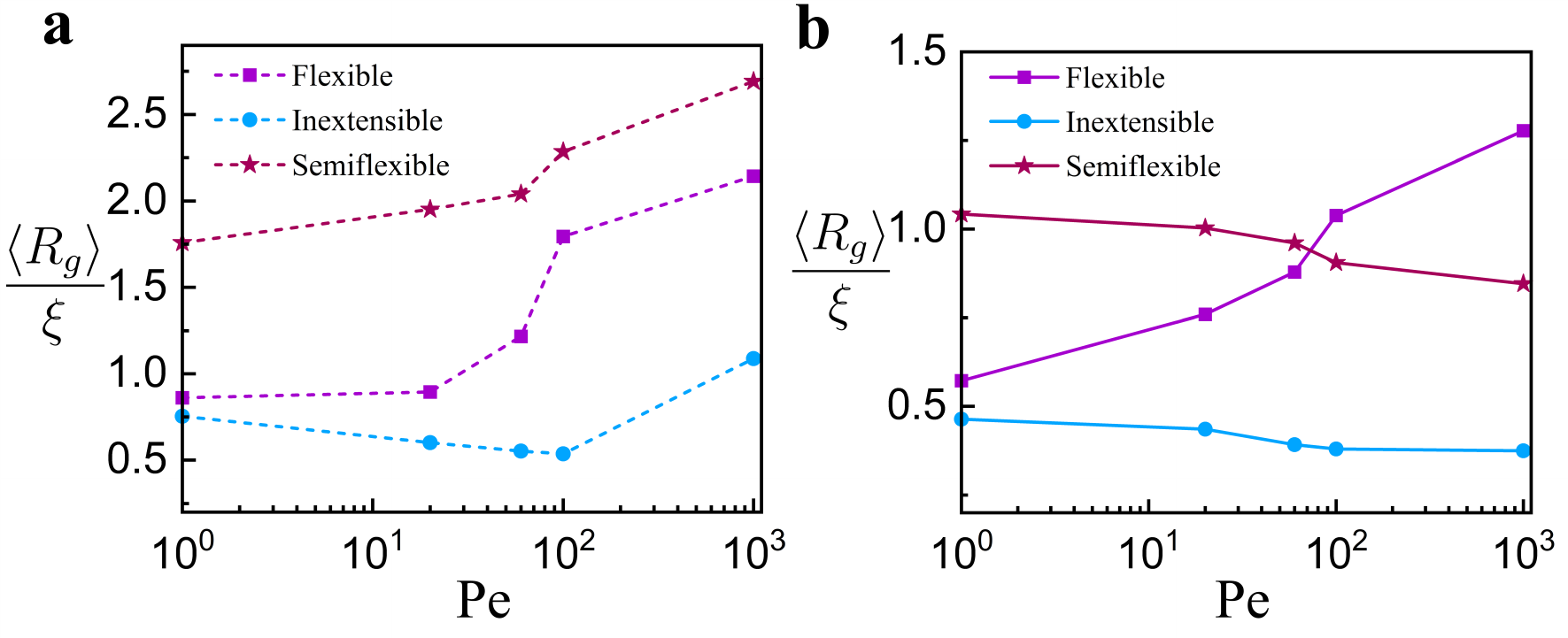}
\caption{\small $\frac{\left < R_g\right >}{\xi}$ $vs$ $\text{Pe}$ for (a) linear chains and (b) rings subjected to different $\text{Pe}$ in porous medium with $\xi = 6.95$.}\label{fig:avg_size}
\end{figure*}

\noindent Next, to elucidate the physical origin of distinctly different migration modes of linear chains and rings in porous media, we analyze the conformations attained by them in porous media and in the unconfined space. For this purpose, we perform gyration tensor analysis (see Supplementary Material for details) and generate the probability distribution of $R_g$, $\text{P}(R_g)$ for a range of activities ($\text{Pe}$). In unconfined media, the flexible and inextensible linear chains swell with increasing $\text{Pe}$, and hence the most probable $R_g$ shifts to larger $R_g$ values (insets of \textbf{Fig.~\ref{fig:conformation}(a, c)}). Semiflexible linear chains shrink at lower activity, and the probability density profile broadens further as the linear chains swell with increasing activity (inset of \textbf{Fig.~\ref{fig:conformation}e}), which is consistent with the previous observations~\cite{eisenstecken2017internal}. In unconfined space, flexible rings also swell with increasing $\text{Pe}$ like the flexible linear chains (inset of \textbf{Fig.~\ref{fig:conformation}b}), while for inextensible and semiflexible rings even though there is an enhancement in conformational fluctuations, but the most probable $R_g$ remains the same (insets of \textbf{Fig.~\ref{fig:conformation}(d, f)}). However, the linear chains possess a larger size than the rings made of the same number of monomers (\textbf{Fig.~\ref{fig:avg_size}}). Swelling of the linear chains and rings is caused by activity, which pushes the monomers away, but the higher spring constant of the inextensible and high bending rigidity of semiflexible rings restrict conformational fluctuations. Therefore, their most probable $R_g$ values are almost independent of $\text{Pe}$ for the range of activities considered (insets of \textbf{Fig.~\ref{fig:conformation}(d, f)}).\\

\noindent Active linear chains and rings depict interesting conformational characteristics depending on their structure in porous media, unlike in unconfined space. Flexible linear chains swell with increasing $\text{Pe}$ (\textbf{Fig.~\ref{fig:conformation}a}) with $\left < R_g\right >$ smaller compared to the unconfined space due to pore confinements (\textbf{Fig.~\ref{fig:avg_size}a}). Inextensible linear chains shrink up to moderate activities followed by swelling at very high $\text{Pe}$ (\textbf{Fig.~\ref{fig:conformation}c}). On the other hand, semiflexible linear chains shrink at very low activity and then swell in the porous media with increasing $\text{Pe}$ (\textbf{Fig.~\ref{fig:conformation}e}). In porous media, flexible rings also swell with activity like linear ones. Inextensible rings shrink at lower and moderate activities and swell at very high $\text{Pe}$. For the semiflexible rings, swelling ceases completely, and the peaks of $\text{P}(R_g)$ shifted to smaller $R_g$ values signifying activity-induced shrinking in porous media (\textbf{Fig.~\ref{fig:conformation}(b, d, f)} and \textbf{Fig.~\ref{fig:avg_size}b}). In the pore confinement, the rings with higher activity more frequently collide with the obstacles with a larger effective force. This subsequently generates more fluctuations along the inward transverse direction of the contour responsible for the shrinking of rings in the pore confinements~\cite{theeyancheri2022migration}. However, such inward transverse fluctuations are largely absent for linear chains due to the extended rod-like structure that helps them to escape from the pore confinements. In porous media, flexible linear chains exhibit a size comparable to or larger than the average pore size $\Big (\frac{\left < R_g\right >}{\xi} \gtrsim 1\Big )$. In contrast, inextensible active linear chains have a relatively smaller size compared to the average pore size $\Big (\frac{\left < R_g\right >}{\xi} < 1\Big )$, while semiflexible linear chains consistently possess a size larger than the average pore size  (\textbf{Fig.~\ref{fig:avg_size}a}). Furthermore, passive flexible rings, due to their circular structure, initially have a size smaller than the average pore size $\Big (\frac{\left < R_g\right >}{\xi} < 1\Big)$. However, with increasing activity, these rings swell and eventually reach a size comparable to or larger than the average pore size $\Big (\frac{\left < R_g\right >}{\xi} \gtrsim 1\Big )$. Intextensible rings behave similarly to linear chains, as their size is smaller than $\xi$. As for semiflexible rings, they initially possess a size larger than $\xi$, but with increasing activity, they shrink, and their size becomes slightly smaller or comparable to the average pore size (\textbf{Fig.~\ref{fig:avg_size}b)}. We also plot the $R_g$ autocorrelation functions for the linear chains and rings, which further support the activity-induced swelling and shrinking of linear chains and rings observed in $\text{P}(R_g)$ in porous media (Fig.~S7). Our analyses unravel the importance of topology in the migration modes of active linear chains and rings in porous media.\\
\begin{figure*}
\centering
\includegraphics[width=0.85\linewidth]{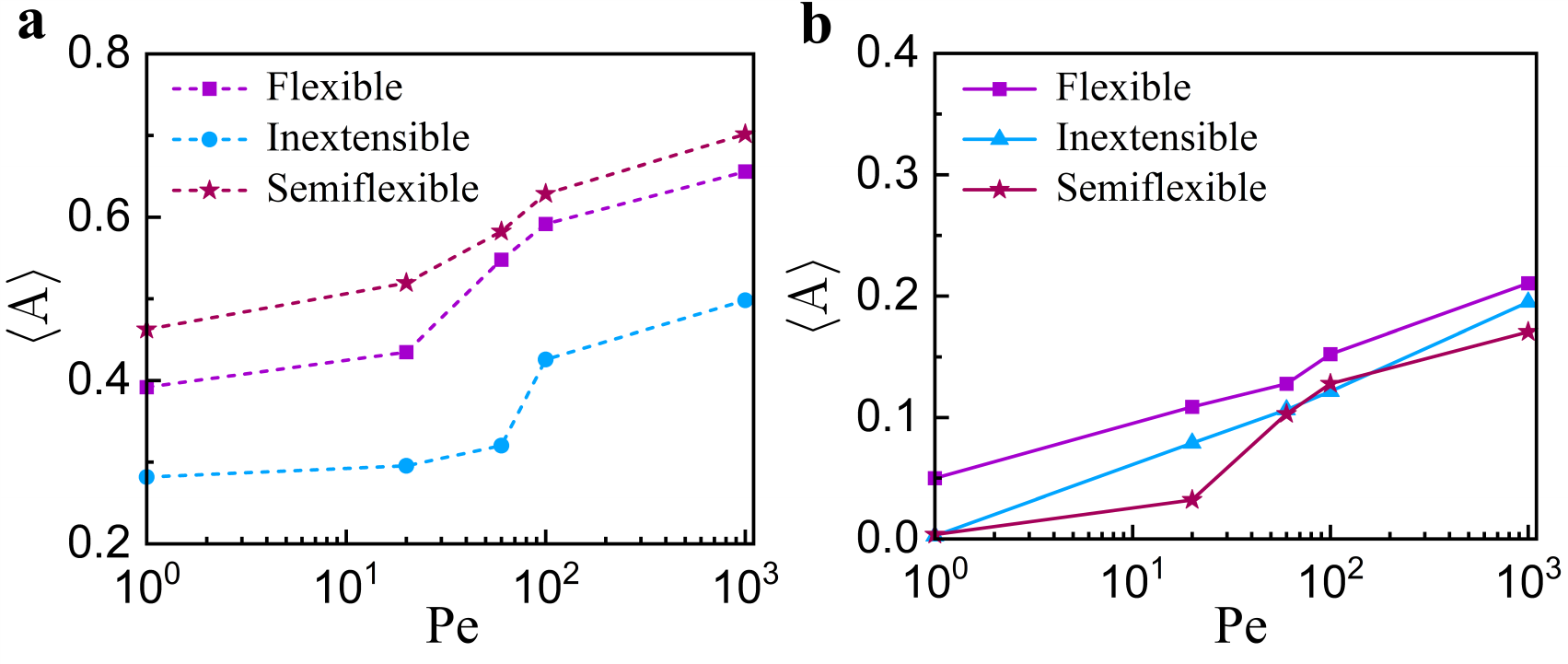}
\caption{\small Average asphericity parameter $\left < \text{A} \right >$ $vs$ $\text{Pe}$ for (a) linear chains (dashed lines) and (b) rings (solid lines) subjected to different activity in the porous medium with $\xi = 6.95$.}\label{fig:asphericity}
\end{figure*}

\noindent To understand how the linear chains and rings deform in the presence of the activity and pore confinements, we compute the shape descriptor asphericity parameter defined as:
\begin{equation} 
 \text{A} = \frac{(\lambda_2 - \lambda_1)^2}{(\lambda_1 + \lambda_2)^2}
\label{eq:asphericity_c4}
\end{equation}
\noindent where $\lambda_1 \, \text{and} \, \lambda_2$ are the eigenvalues of gyration tensor. $\left < \text{A} \right >$ ranges between 0 and 1, where 0 corresponds to circular/collapsed, and 1 corresponds to most extended geometry. $\left < \text{A} \right >$ increases with $\text{Pe}$ for the flexible, inextensible, and semiflexible linear chains and rings (\textbf{Fig.~\ref{fig:asphericity}}). However, semiflexible linear chains possess higher values of $\left < \text{A} \right >$ compared to the flexible linear chains because the semiflexible linear chains display a more extended geometry compared to the flexible ones (\textbf{Fig.~\ref{fig:asphericity}a}). Inextensible linear chains show a shrinking-like behavior at lower activities as reflected by the lower values of $\left < \text{A} \right >$ (\textbf{Fig.~\ref{fig:asphericity}a}). On the contrary, flexible rings have higher values of $\left < \text{A} \right >$ compared to the semiflexible ring because the semiflexible rings prefer to be in the circular geometry, and the flexible rings take elongated structure due to swelling (\textbf{Fig.~\ref{fig:asphericity}b}). This further supports the transient trapping observed for the semiflexible rings. The distributions of the asphericity parameter, $\text{P}(A)$, manifest that linear chains and rings undergo strong deformation with increasing $\text{Pe}$ (Fig.~S8). The extent of straightening is more pronounced for linear chains than the rings owing to the structural restrictions. \\

\noindent We compare the dynamical and conformational properties of flexible and semiflexible linear chains with rings of comparable size $\left (\left <\text{R}_g\right >\right)$, but with the different number of monomers (Fig.~S9 and Fig.~S10). We find that the qualitative trends remain the same as the flexible or semiflexible active linear chains move faster than the rings, and for the semiflexible case, there is a more pronounced difference between their motion in porous media (Fig.~S9). This indicates that apart from the size, polymer topology also plays a crucial role in the porous media. Subsequently, we analyze the conformations of these polymers, which uncover the swelling of flexible linear chains and rings, whereas the semiflexible linear chains behave differently from rings (Fig.~S10). Semiflexible linear chains exhibit activity-induced shrinking followed by swelling, while rings with a size comparable to linear chains shrink with activity in porous media (Fig.~S10). Therefore, the topology-guided dynamics and conformations of polymers with comparable sizes play a pivotal role in their efficient migration in porous environments.

\section{Summary}

\noindent In summary, we present how the motion of active Brownian linear chains and rings is controlled by pore size, activity, polymer topology, and the nature of the polymers in porous media. The dynamics of COM of the linear chains or rings is enhanced by orders of magnitude with activity and show intermediate superdiffusion, unlike the passive case, irrespective of their topology. Flexible active linear chains and rings migrate smoothly through the pores, and inextensible ones also show a similar trend with increasing activity. On the other hand, the semiflexible linear chains exhibit distinctly different dynamics compared to the semiflexible rings with the same number of monomers in the porous media. Semiflexible linear chains smoothly navigate through the porous media by adopting a straight rod-like structure. In contrast, semiflexible rings display a transition from trapping inside the pore confinements at lower activities due to stretched circular-like conformations to escaping at higher activities facilitated by the enhanced conformational fluctuations. This portrays the effect of topology-facilitated navigation in porous media.\\

\noindent Our results depict that there is a considerable difference in conformational properties of linear chains and rings while exploring the porous media. In porous media, the flexible linear chains and rings both swells with activity. Inextensible linear chains shrink at lower activity and swell at very high activities, while inextensible rings show activity-induced shrinking in porous media. For the semiflexible case, the linear chains shrink at lower activities, followed by swelling at higher activities, but the rings exhibit only shrinking with increasing activity.\\

\noindent The porous architecture also plays a significant role as the diffusion largely depends on the pore size. A larger pore size facilitates the motion of the rings, while a pore size smaller than the ring size accelerates the dynamics of the linear chains compared to the rings. Our findings disclose that pore confinement and the topology of the active agents substantially alter the dynamical and conformational properties of active polymers. The physics underlying the phenomena we report here relies on the transport of active linear chains and rings through planar disordered porous environment driven by the combined effects of confinement and activity. Hence, further studies on the migration of the active linear chains and rings in three-dimensional porous environments are anticipated in the future as the 3D porous architecture becomes more complex, and thus the structure of linear and ring polymers will no longer be identical, which could lead to qualitative changes in their behavior.\\  

\section*{Acknowledgments}
\noindent L.T. thanks UGC for a fellowship. S.C. thanks DST Inspire for a fellowship. R.C. acknowledges SERB for funding (Project No. MTR/2020/000230 under MATRICS scheme). T.B. acknowledges NCBS-TIFR for research funding. We acknowledge the SpaceTime-2 supercomputing facility at IIT Bombay for the computing time.\\

\clearpage
\onecolumngrid \section*{Supplementary Material}
\renewcommand{\thefigure}{S\arabic{figure}}
\setcounter{figure}{0}
\renewcommand{\thesubsection}{\roman{subsection}}
\setcounter{subsection}{0}
\subsection{Gyration Tensor Analysis}

\noindent To evaluate the shape fluctuations or how the linear chains and rings deform, we calculate the gyration tensor of the conformations defined as,
 \begin{equation}
S = \left (\begin{smallmatrix}
 \sum_i (x_i-x_\text{com})^2 & \sum_i (x_i-x_\text{com})(y_i-y_\text{com}) \vspace{4mm} \\ \sum_i (x_i-x_\text{com})(y_i-y_\text{com}) & \sum_i (y_i-y_\text{com})^2 
\end{smallmatrix}\right )
\end{equation}

\noindent where, $x_\text{com} \, \text{and} \, y_\text{com}$ represent the x and y components of the COM position respectively. Further, we compute the eigenvalues $\lambda_1 \, \text{and} \, \lambda_2$ of the gyration tensor by diagonalizing the matrix S. These eigenvalues are used to define the shape descriptor, asphericity parameter, $\text{A} = \frac{(\lambda_2 - \lambda_1)^2}{(\lambda_1 + \lambda_2)^2}$. It measures the deviation from the spherical symmetry. The asphericity parameter is 1 for a perfect rod-like or linear topology and 0 for a circular structure. \\

\noindent We extract the radius of gyration,
\begin{equation}
R_g = \left [ \frac{1}{N} \sum_{i = 1}^N (r_i - r_\text{com})^2\right ]^{\frac{1}{2}}
\end{equation}

\noindent from gyration tensor for each time-step of every simulation after the system reaches the steady state. Here $r_\text{com}$ is the center of mass of the linear chains or rings. Then a single trajectory is created by stitching different individual trajectories together. This single trajectory is binned to construct a histogram from which the ensemble-averaged probability distribution $\text{P}(R_g)$ is obtained. \\

\begin{figure}[h!]
\centering
\includegraphics[width=0.95\linewidth]{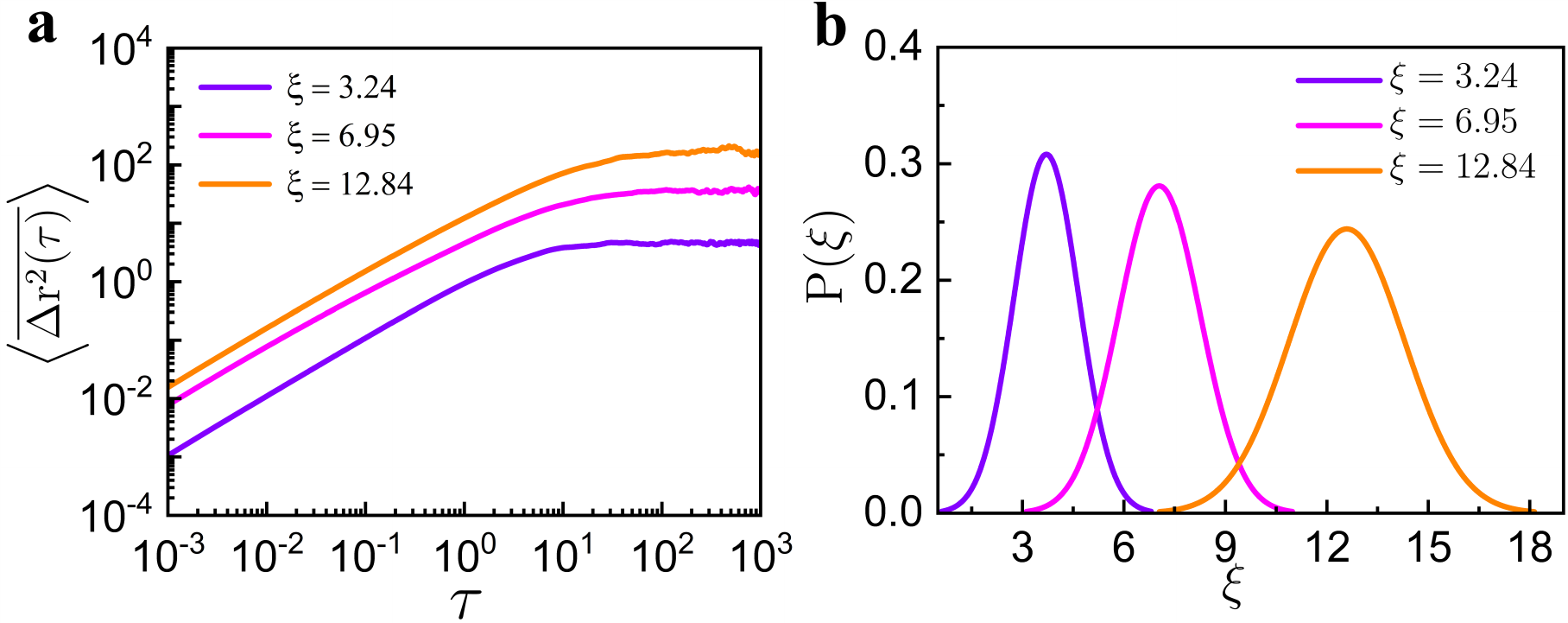}
\caption{(a) Log-log plot of $\left\langle{\overline{\Delta \text{r}_{c}^{2}(\tau)}}\right\rangle$ $vs$ $\tau$ of the passive tracer particle in different porous media. (b) Distribution of pore spaces, $\text{P}(\xi)$ for different porous media.}\label{fig:pore_size}
\end{figure}
\begin{figure*}
\centering
\includegraphics[width=0.95\linewidth]{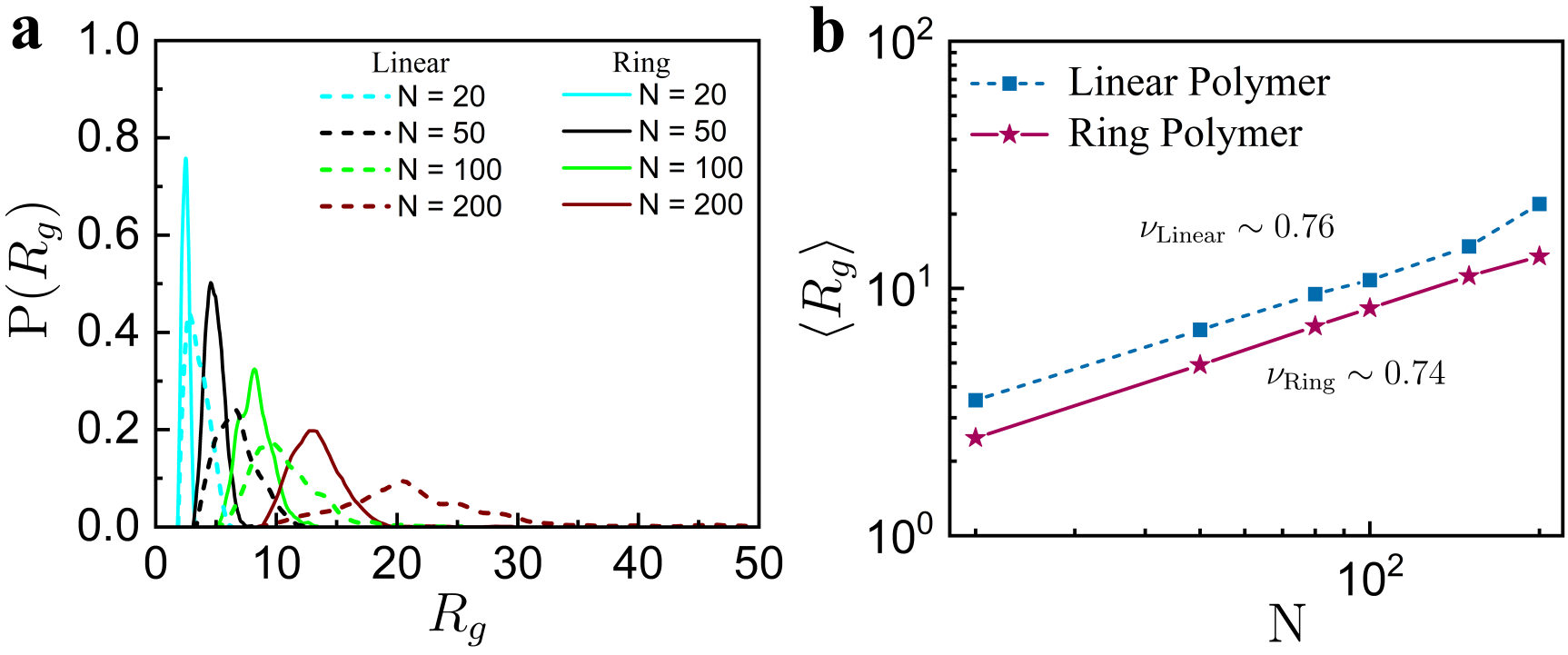}
\caption{(a) $\text{P}(R_g)$ $vs$ $R_g$ and (b) $\left < R_g \right >$ $vs$ $N$ for flexible linear chains (dashed lines) and rings (solid lines) of different monomers N = 20, 50, 80, 100, 150, 200 in unconfined space. Linear fitting gives the Flory exponents $\nu_{\text{Linear}}$ and $\nu_{\text{Ring}}$.}\label{fig:scaling_N}
\end{figure*}
\begin{figure*}
\centering
\includegraphics[width=0.95\linewidth]{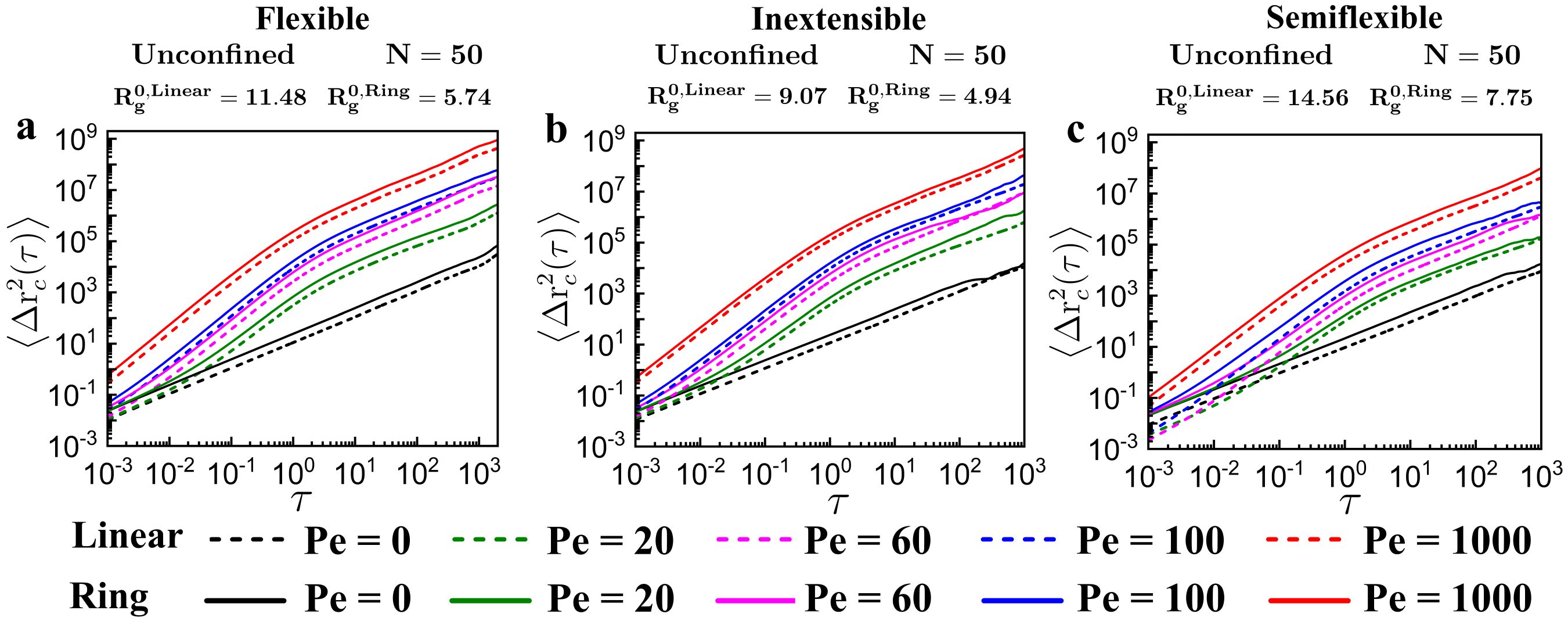}
\caption{Log-log plot of $\left\langle{\overline{\Delta \text{r}_{c}^{2}(\tau)}}\right\rangle$ $vs$ $\tau$ for (a) flexible, (b) inextensible, (c) semiflexible linear chains (dashed lines) and rings (solid lines) (N = 50) subjected to different activity in unconfined media.}\label{fig:msd_free}
\end{figure*}
\begin{figure*}
\centering
\includegraphics[width=0.95\linewidth]{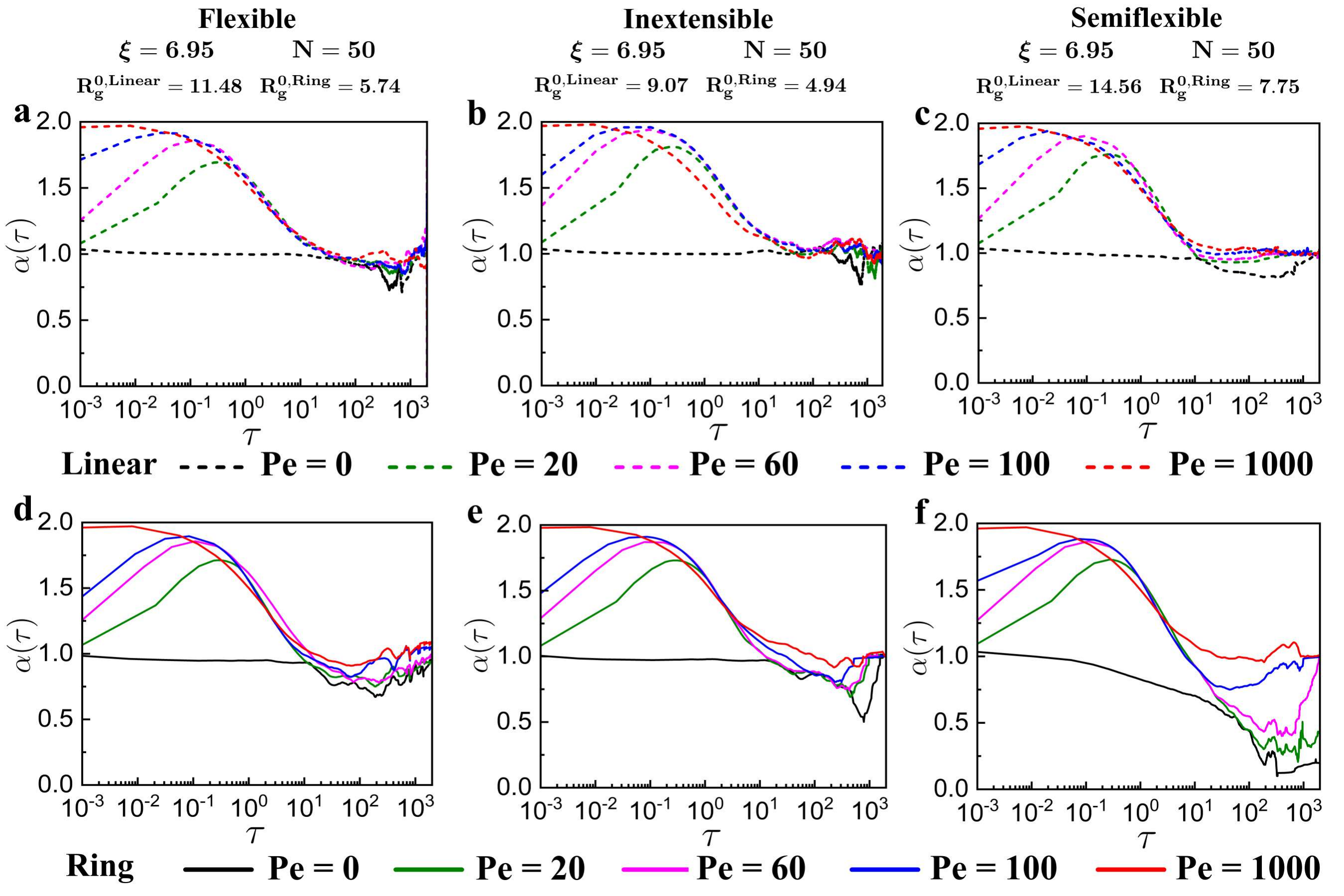}
\caption{(a) $\alpha(\tau)$ $vs$ $\tau$ for the flexible, inextensible, semiflexible linear chains (a, b, and c respectively) and rings (d, e, and f respectively) in porous media with $\xi = 6.95$.}\label{fig:exponent_Fa}
\end{figure*}
\begin{figure*} 
\centering
\includegraphics[width=0.95\linewidth]{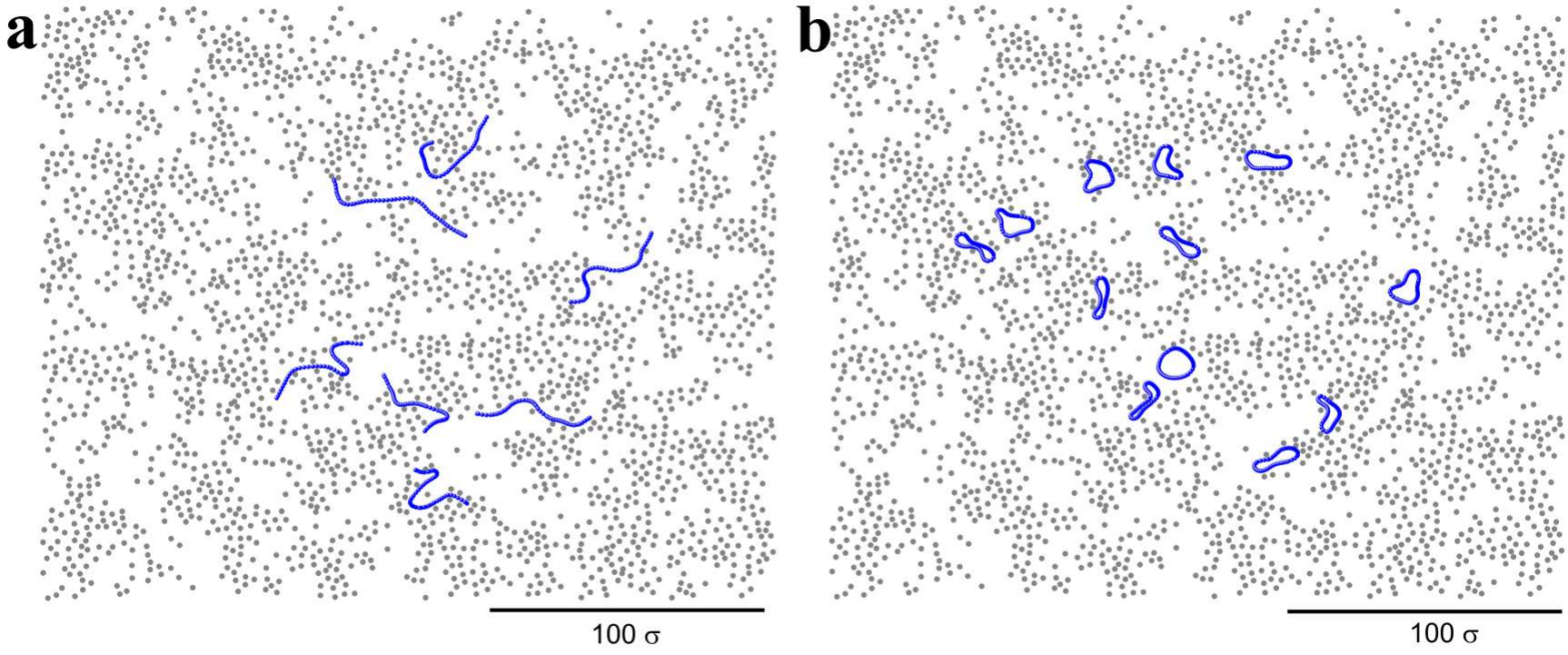}
\caption{(a) Straight/rod-like conformations of active ($\text{Pe} = 60$) semiflexible linear chains and (b) stretched conformations of active semiflexible rings in porous media ($\xi = 6.95$). The conformations are extracted from the most probable values of corresponding $\text{P}(R_g)$.}\label{fig:conformation}
\end{figure*}
\begin{figure*}
\centering
\includegraphics[width=0.95\linewidth]{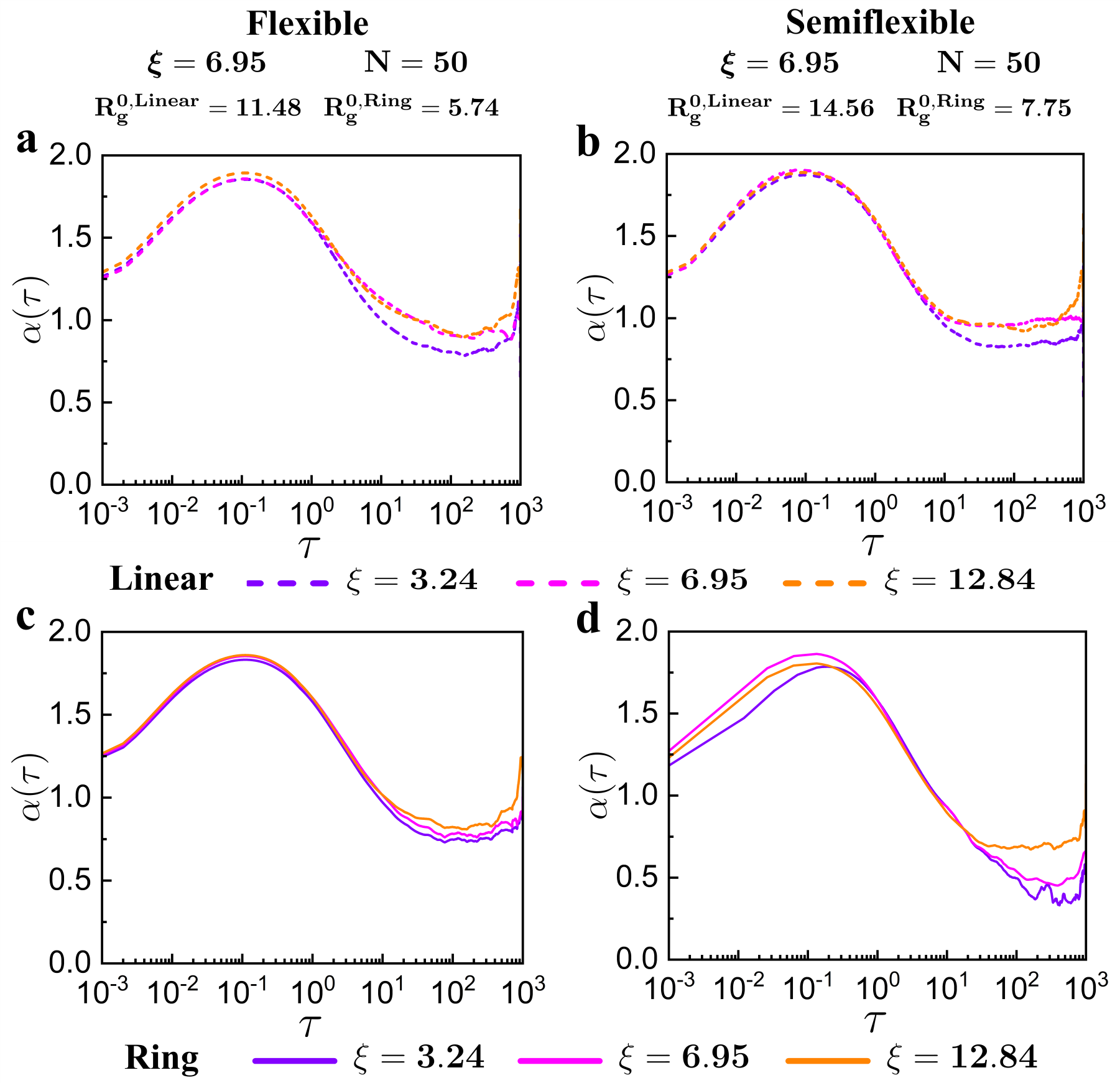}
\caption{(a) $\alpha(\tau)$ $vs$ $\tau$ for the flexible and semiflexible active ($\text{Pe} = 60$) linear chains (a and b) and rings (c and d) (N = 50) in porous media for different $\xi = 3.24, 6.95, \text{and} 12.84$.}\label{fig:exponent_xi}
\end{figure*}
\begin{figure*}
\centering
\includegraphics[width=0.95\linewidth]{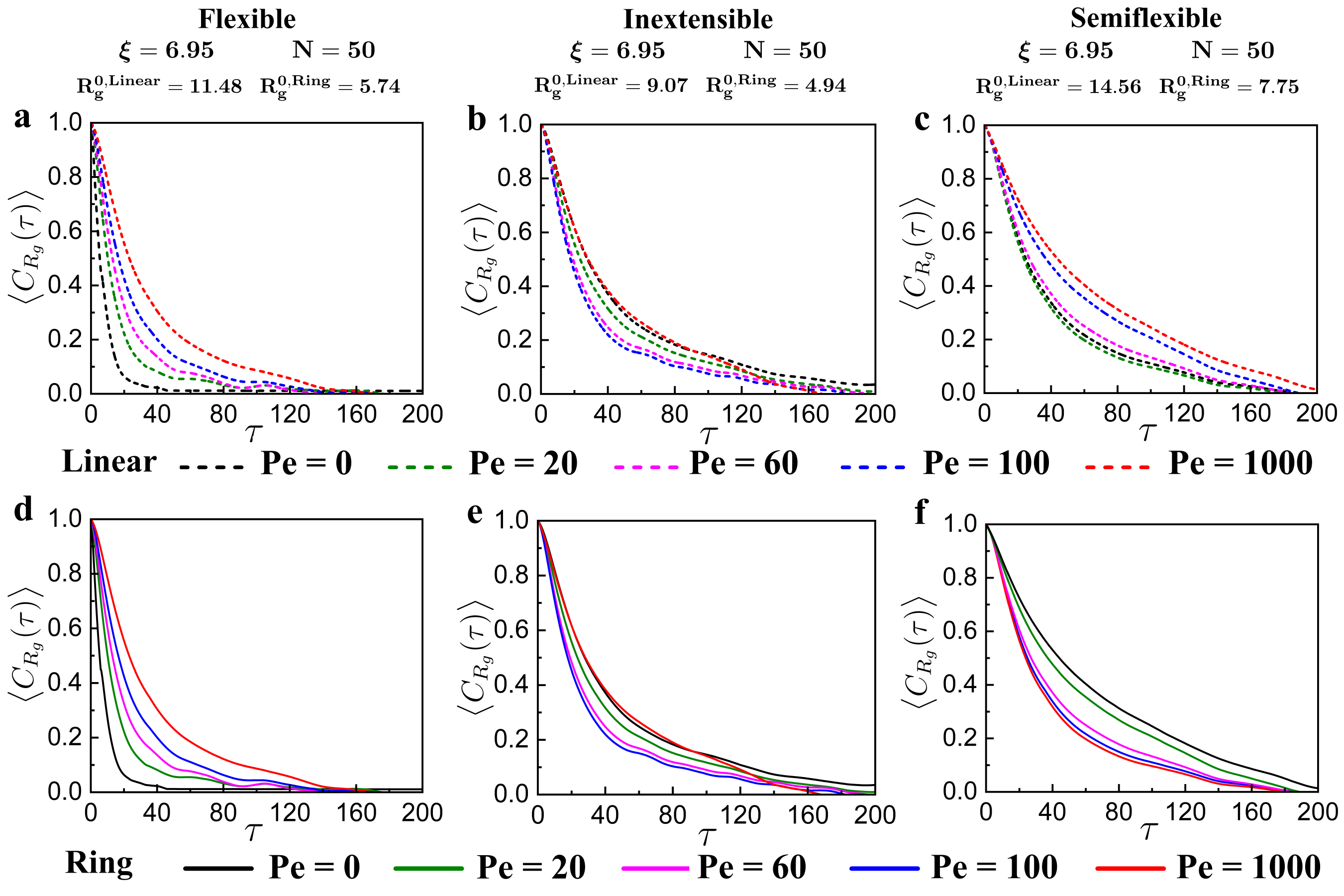}
\caption{(a) $\left < C_{R_g}(\tau)\right>$ $vs$ $\tau$ for flexible, inextensible, semiflexible linear chains (a, b, and c respectively) and rings (d, e, and f respectively) (N = 50) subjected to different activity in the porous medium with $\xi = 6.95$.}\label{fig:ACF}
\end{figure*}
\begin{figure*}
\centering
\includegraphics[width=0.95\linewidth]{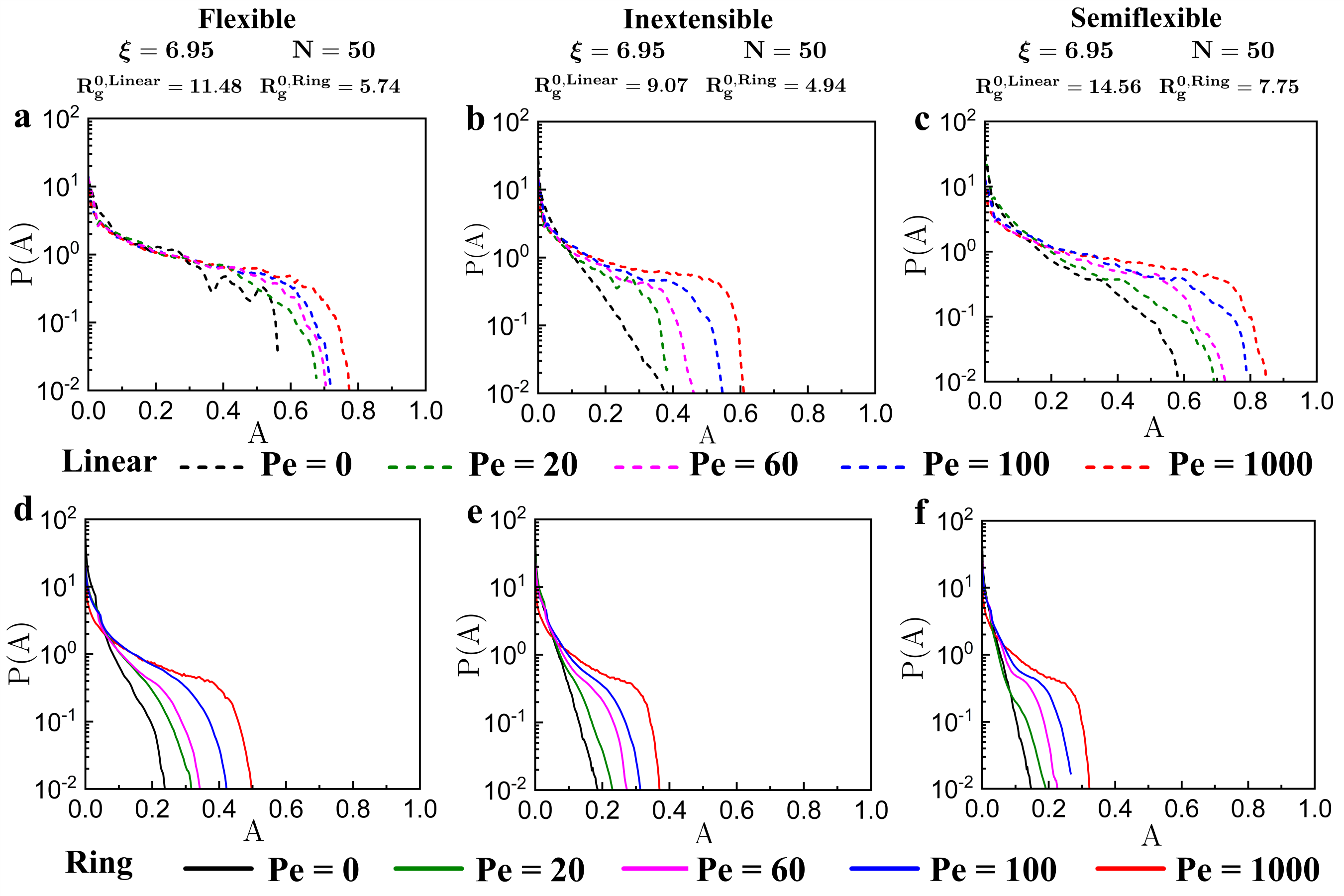}
\caption{$\text{P}(\text{A})$ $vs$ $\text{A}$ for flexible, inextensible, semiflexible linear chains (a, b, and c respectively) and rings (d, e, and f respectively) (N = 50) subjected to different activity in the porous medium with $\xi = 6.95$.}\label{fig:asph_dist}
\end{figure*}
\begin{figure*}
\centering
\includegraphics[width=0.9\linewidth]{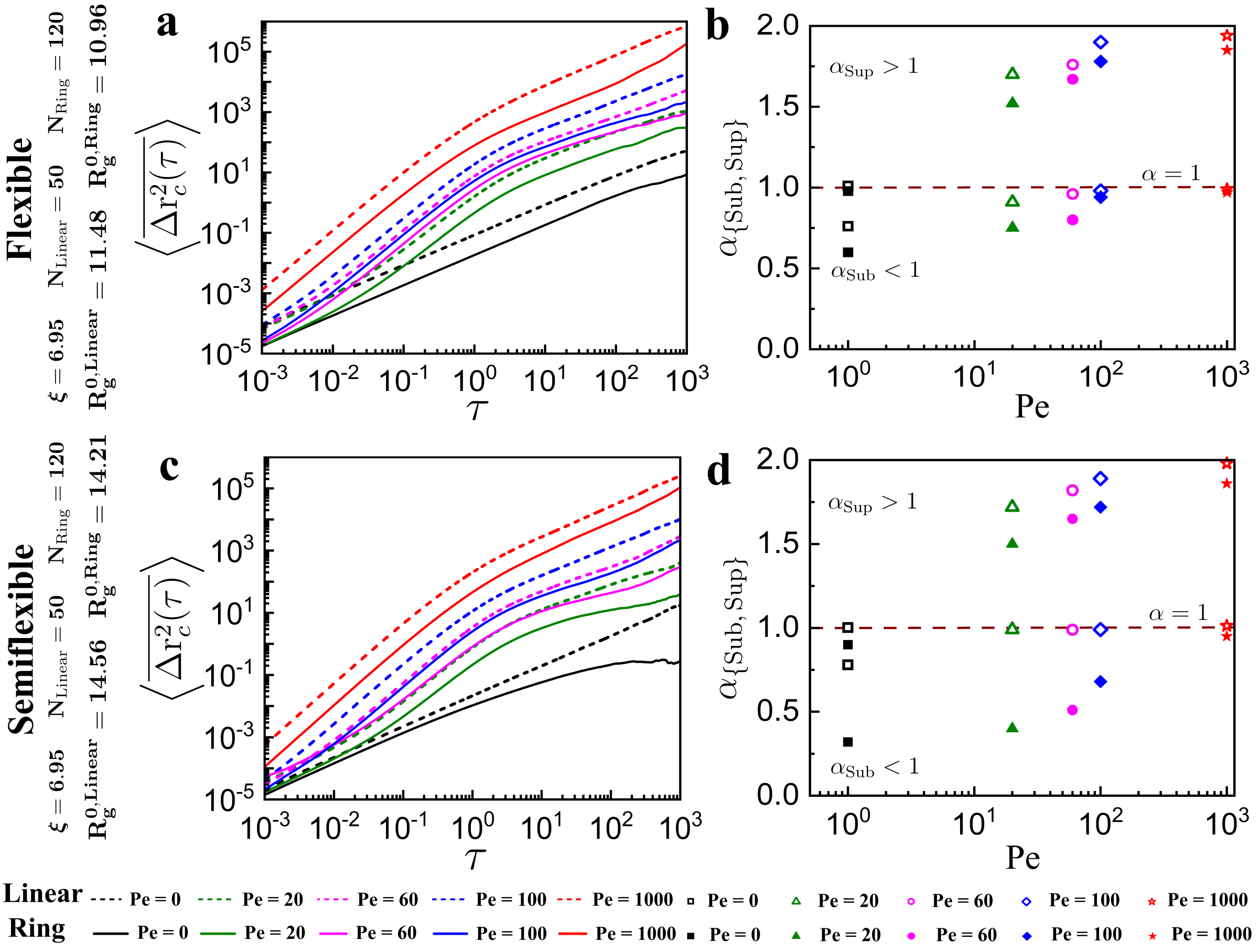}
\caption{\small $\left\langle{\overline{\Delta \text{r}_{c}^{2}(\tau)}}\right\rangle$ $vs$ $\tau$ and $\alpha_{\{\text{Sub, Sup}\}}$ $vs$ $\text{Pe}$ for flexible (a, b), and semiflexible (c, d) linear chains (dashed lines or open symbols) and rings (solid lines or solid symbols) subjected to different activity in the porous medium with $\xi = 6.95$. $\alpha_{\{\text{Sub, Sup}\}}$ illustrates the subdiffusion ($\alpha_{\text{Sub}} < 1$) and superdiffusion ($\alpha_{\text{Sup}} > 1$) exponents, respectively. Brown dashed line represents $\alpha = 1$. Here, the rings are comparable in size to the linear chains by considering different numbers of monomers ($N_{\text{Ring}} = 120$ and $N_{\text{Linear}} = 50$).}\label{fig:dynamics_Rg_same}
\end{figure*}
\begin{figure*}
\centering
\includegraphics[width=0.9\linewidth]{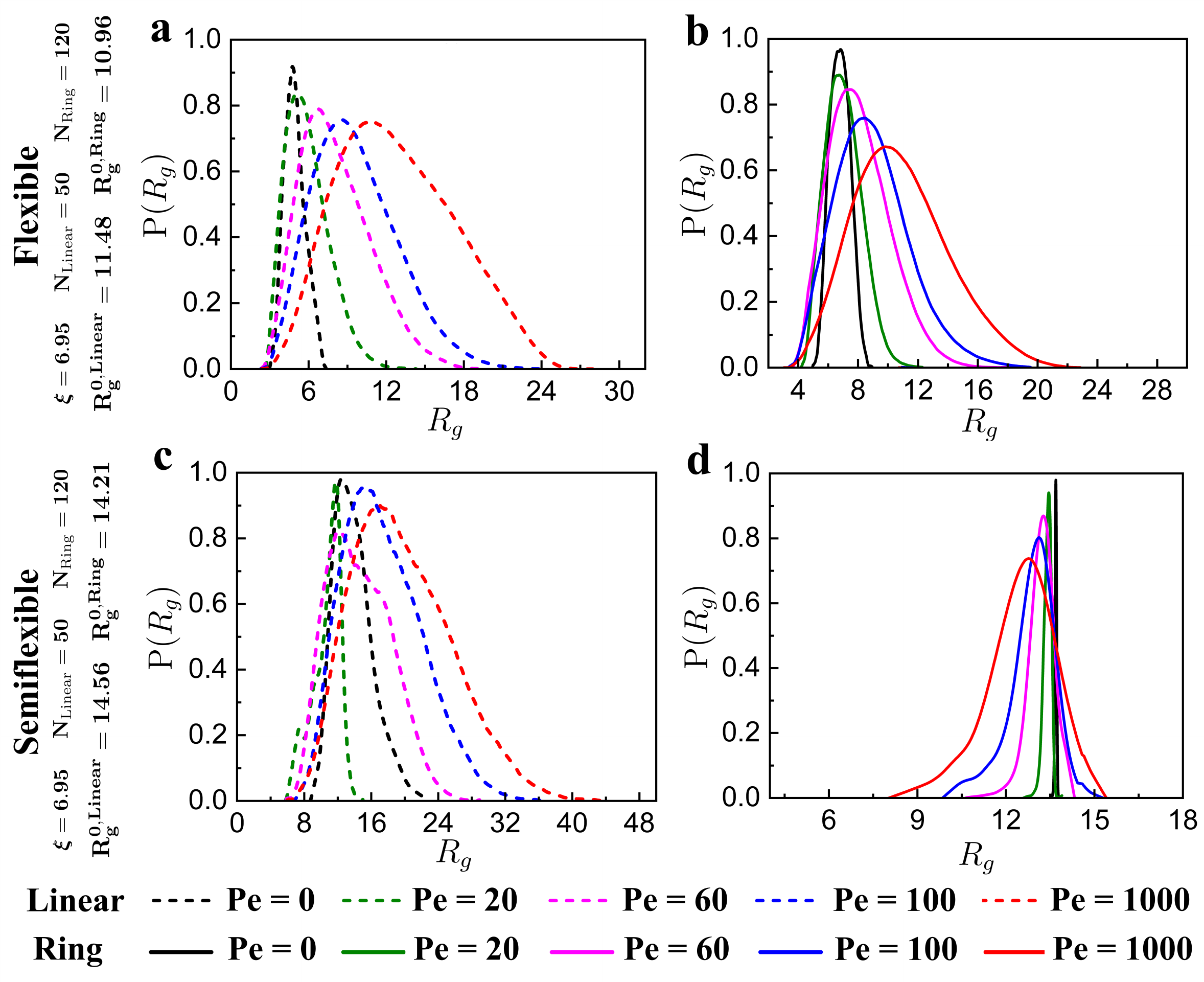}
\caption{\small $\text{P}(R_g)$ $vs$ $R_g$ for flexible (a, b), and semiflexible (c, d) linear chains (dashed lines) and rings (solid lines) with comparable $<\text{R}_g>$ subjected to different activity in the porous medium with $\xi = 6.95$. Here, the rings are comparable in size to the linear chains by considering different numbers of monomers ($N_{\text{Ring}} = 120$ and $N_{\text{Linear}} = 50$).}\label{fig:conformation_Rg_same}
\end{figure*}

\clearpage
\noindent \textbf{Movie Description} \\

\begin{enumerate}
\item \textbf{Movie\_S1} \\
\noindent The motion of flexible active ($\text{Pe} = 60$) linear chain in random porous media ($\xi= 6.95$). The linear chain undergoes conformational changes while migrating through the pore confinements and results activity-induced swelling in the porous media. \\

\item \textbf{Movie\_S2} \\
\noindent The motion of flexible active ($\text{Pe} = 60$) ring in random porous media ($\xi= 6.95$). The ring undergoes a series of conformational changes while migrating through the pore confinements in the media. \\

\item \textbf{Movie\_S3} \\
\noindent The motion of inextensible active ($\text{Pe} = 60$) linear chain in random porous media ($\xi= 6.95$). The smaller size and linear topology help the linear chain to smoothly move through the pore confinements.\\

\item \textbf{Movie\_S4} \\
\noindent The motion of inextensible active ($\text{Pe} = 60$) ring in random porous media ($\xi= 6.95$). The smaller size of the ring facilitates smooth and faster migration through the pore spaces compared to flexible (Movie\_S2) and semiflexible (Movie\_S6) rings. \\ 

\item \textbf{Movie\_S5} \\
\noindent The motion of semiflexible active ($\text{Pe} = 60$) linear chain in random porous media ($\xi= 6.95$). The semiflexible linear chain prefers to be in an extended conformation, and thus it shows activity-induced swelling in the porous media. The extended structure of the linear chain facilitates the movement by occupying multiple pore confinements without getting trapped, which is absent for the ring analogue (Movie\_S6). \\

\item \textbf{Movie\_S6} \\
\noindent The motion of semiflexible active ($\text{Pe} = 60$) ring in random porous media ($\xi= 6.95$). The average size of the semiflexible ring is larger than flexible and inextensible rings. The higher bending rigidity of the ring restricts conformational fluctuations, and thus it gets trapped in the pore confinements. The activity-induced shrinking of the ring helps in escaping from the traps and migrating through the porous media.\\   
\end{enumerate}
\clearpage
%
\end{document}